\documentclass[aps,twocolumn,draft]{revtex4}
\input epsf
\begin{document}

\title{
Competition between different order parameters in a quasi-one-dimensional
superconductor
}
\author{A.V. Rozhkov}

\affiliation{
Institute for Theoretical and Applied Electrodynamics RAS,
Moscow, ul. Izhorskaya 13/19, 125412, Russian Federation
}


\begin{abstract}
We show that, under rather general assumptions, the phase diagram of a
quasi-one-dimensional repulsive Fermi system
consists of two ordered phases: the density wave, spin or charge, and the
superconductivity. It is demonstrated that the symmetry of the
superconducting order parameter is a non-universal property sensitive to
microscopic details of the model. Three potentially stable superconducting
states are identified: they are triplet $f$-wave, singlet
$d_{x^2-y^2}$-wave,
and
$d_{xy}$-wave.
Presence of multiple competing superconducting states implies that for a
real material this symmetry is difficult to predict theoretically and hard to
probe experimentally, since artifacts of theoretical approximations or
variations in experimental conditions could tip the balance between the
superconducting phases.
\end{abstract}
\date{\today}
\maketitle
\hfill

\section{Introduction}
Many theorists hold that the superconductivity in quasi-one-dimensional
(Q1D) materials may be explained with the help non-phonon many-body
mechanism
\cite{mech,guay,rozhkov,rozhkov_cond-mat,nickelII,dup,nickel,
review_RG1,review_RG2,fuseyaI,fuseyaII,tanaka,kuroki,flex,review_RPA,
belmechri,belmechriII}.
Yet, there is substantial disagreement about the details of such
mechanism. Moreover, different theories predict different order parameter
symmetries:
$p-$,
$d_{x^2 - y^2}$-,
$f$-, or
$d_{xy}$-wave
\cite{review_RG1,review_RG2,review_RPA,fuseyaI}.
The experimental findings
\cite{NMR,NMR2,imp,field,field2,no_nodes,knight,universal}
have been unable to resolve these controversies.

In this article we explain why establishing the symmetry of the order
parameter is such a hard task. It will be demonstrated that within the
framework of the non-phonon mechanism there are three metastable
superconducting states:
$d_{x^2 - y^2}$-, $f$-, and $d_{xy}$-wave,
which compete to become the ground state.
This circumstance has an important implication for both theory and
experiment: to find the ground state symmetry one has to resolve small
energy differences between the contesting states. Thus, an approximate
theoretical scheme might calculate these differences completely wrong; a
slight variation of the model may cause a phase transition. An
experimentalist has to keep in mind that, despite the obvious similarities
between various families of the Q1D superconducting
materials in chemistry and structure, despite ``universality" of their
phase diagram \cite{universal}, the superconducting order does not
have to be identical in all these metals. Moreover, even an individual
sample may experience phase transitions between different types of
superconductivity as the pressure or magnetic field change (e.g.,
\cite{knight}).
This suggests that experimental detection of the order parameter symmetry
should be done for a specific material, rather than a class of materials;
a reliable theoretical prediction of this symmetry is nearly impossible.

Let us briefly outline main ideas of the discussion. Due to presence of
strong many-body effects,
the Q1D metal cannot be treated by the usual mean-field theory. Thus, to
circumvent this difficulty, we derive the low-energy effective Hamiltonian.
Unlike the original microscopic model, the anisotropy of the effective
model
is low, and it may be studied with the help of the mean-field
approximation.
Application of mean-field theory to the effective model
\cite{mech,guay}
allows us to find the phase diagram of the Q1D metal. At good nesting the
system freezes into a spin or charge density wave. The antinesting destroys
the density wave phase and makes superconductivity possible. In
superconducting phase we find three order parameters, which may be stable in
our model.

The paper is organized as follows. The Q1D model and its effective
low-energy description is presented in Sect.~\ref{model}. 
The phase diagram is obtained in Sect.~\ref{phase_diag}.
Sect.~\ref{discussion} is reserved for discussion of the results.

\section{Q1D model and its effective Hamiltonian}
\label{model}

\subsection{Bare Hamiltonian}

The model we study is well-known. The system consists of 1D chains, which
are arranged into a square array to form a three-dimensional (3D) system.
Its Hamiltonian has the form:
\begin{eqnarray}
H = \sum_i H_i^{\rm 1D} 
+ 
\sum_{ij} H^{\rm hop}_{ ij} + H^{\rho \rho}_{ ij},
\label{Hbare}
\end{eqnarray}
where $H_i^{\rm 1D}$ is the Hamiltonian of an individual 1D chain $i$.
\begin{eqnarray} 
{H}_i^{\rm 1D}
&=&
H^{\rm kin}_i 
+ 
H^{\rho\rho}_i,
\label{H1D}
\\
{H}_i^{\rm kin}
&=&
-{i} v_{\rm F} \sum_{p\sigma}
p \int dx 
\/\/ \colon\!
        \psi^\dagger_{p\sigma i} 
        (
             \nabla
             \psi^{\vphantom{\dagger}}_{p \sigma i}
        ) 
  \colon,
\label{Hkin}
\\
H^{\rho\rho}_i
&=&
\int dx 
\Big[
	- 
	J_{2k_{\rm F}}
	{\bf S}_{{2k_{\rm F}}i}
	\cdot 
	{\bf S}_{{ - 2k_{\rm F}}i} 
\label{rhorho_exp}
\\
\nonumber 
	&-& 
	g_{2k_{\rm F}}
	\rho_{{ 2k_{\rm F}}i}
	\rho_{{ -2k_{\rm F}}i} 
	+
	g_{4} 	\left( 
			\rho_{{\rm L}\uparrow i} 
			\rho_{{\rm L}\downarrow i} 
       	        	+ 
       	        	\rho_{{\rm R}\uparrow i} 
       	        	\rho_{{\rm R}\downarrow i}  
       		\right) 
\Big].
\nonumber 
\\
\rho_{p \sigma}&=&\colon 
			\psi^{\dagger}_{p \sigma}
			\psi^{\vphantom{\dagger}}_{p \sigma}
		  \colon,
\\
\rho_{2 k_{\rm F}}&=&\sum_\sigma	
			\psi^{\dagger}_{{\rm R} \sigma}
			\psi^{\vphantom{\dagger}}_{{\rm L} \sigma},
\quad
\rho_{-2 k_{\rm F}}^{\vphantom{\dagger}}  = 
\rho_{2 k_{\rm F}}^\dagger,
\\
{\bf S}_{2k_{\rm F} }
&=&
\sum_{\sigma \sigma'} \vec{\tau}_{\sigma \sigma'} 
\psi^{\dagger}_{{\rm R} \sigma }
\psi^{\vphantom{\dagger}}_{{\rm L} \sigma' } ,
\quad
{\bf S}_{-2k_{\rm F}}^{\vphantom{\dagger}}
= 
{\bf S}_{2k_{\rm F}}^{\dagger}.
\end{eqnarray}
Here index
$p = \pm 1$
labels different chiralities of 1D electrons,
right-movers 
$\psi_{{\rm R}\sigma}$ 
($p = 1$) and left-movers 
$\psi_{{\rm L}\sigma}$ 
($p=-1$). Vector 
$\vec{\tau}$
is composed of three Pauli matrices. The coupling constants 
$g_{4, {2 k_{\rm F}}}$ 
and 
$J_{{2 k_{\rm F}}}$
are positive, which corresponds to repulsion between electrons. The model's
microscopic cutoff is denoted by $\Lambda$.

We expressed our Hamiltonian $H^{\rho\rho}_i$ is a somewhat unusual form.
In a more traditional notation this operator looks as such:
\begin{eqnarray}
{H}^{\rho\rho}_i
&=&
\int dx 
\Big[
	g_1  \rho_{2k_{\rm F} i}   \rho_{-2k_{\rm F} i}
	+
	g_2 \sum_{\sigma \sigma'} 
	\rho_{{\rm L}\sigma i} 
	\rho_{{\rm R}\sigma' i} 
\label{rhorho}
%
\\
&+&
g_{4} 	\left( 
		\rho_{{\rm L}\uparrow i} 
		\rho_{{\rm L}\downarrow i} 
               	+ 
               	\rho_{{\rm R}\uparrow i} 
               	\rho_{{\rm R}\downarrow i}  
       	\right) 
\Big],
\nonumber 
\\
g_{2k_{\rm F}}&=&\frac{g_2}{2} - g_1,
\label{g2kF}
\\
J_{2k_{\rm F}}&=&\frac{g_2}{2}.
\label{J2kF}
\end{eqnarray}
Both forms are absolutely equivalent. Eq.(\ref{rhorho_exp}) suits us more,
for it explicitly shows the couplings of the density waves.

Closely located chains are coupled by single-electron hopping 
$H^{\rm hop}_{ij}$ and electron-electron interaction $H^{\rho\rho}_{ij}$:
\begin{eqnarray}
{H}_{ij}^{\rm hop}
&=&
- t(i-j) 
\sum_{p \sigma}
\int dx
\left(
        \psi^\dagger_{p\sigma i}
        \psi^{\vphantom{\dagger}}_{p\sigma j} 
        +
        {\rm H.c.}
\right), 
\\
{H}_{ij}^{\rho\rho}
&=&
\int dx
\Big[
	g^\perp_0 (i-j) 
	\rho_{i} \rho_{j} 
\label{perp}
\\
&+&g_{2k_{\rm F}}^\perp (i-j) 
	\left( 
		\rho_{2k_{\rm F} i}
		\rho_{-2k_{\rm F} j} 
		+ 
		{\rm H.c.}
	\right)
\Big],
\nonumber 
\\
\rho&=&\sum_{p \sigma} \rho_{p \sigma}.
\end{eqnarray}
We assume that our microscopic model is characterized by the
following hierarchy of material constants. The anisotropy ratio is small:
\begin{eqnarray}
r = t / v_{\rm F} \Lambda \ll 1.
\label{ratio}
\end{eqnarray}
In addition, the 
$2k_{\rm F}$ 
coupling constants are smaller than the coupling constant corresponding to
interactions of smooth components of the density, and the in-chain
interactions are larger (or much larger) than the inter-chain interactions:
\begin{eqnarray}
g_{2k_{\rm F}}^\perp \lesssim g_0^\perp 
<
g_{2k_{\rm F}} \lesssim J_{2k_{\rm F}} \sim g_4 \ll v_{\rm F}.
\label{2kF<0}
\end{eqnarray}
The smallness of the transverse couplings as compared to the in-chain
coupling constants assures that at high energy the system may be viewed as a
collection of weakly perturbed 1D conductors. The smallness of all coupling
constants as compared to 
$v_{\rm F}$
indicates that weak coupling arguments may be applied.

\subsection{Effective description}

It is tempting to study the low-temperature phase diagram of $H$,
Eq.~(\ref{Hbare}), with the help of the mean field approximation. Yet, one
has to keep in mind that this idea is wrong. It is demonstrated by Prigodin
and Firsov \cite{prigodin_firsov}, who investigate the renormalization group
(RG) flow of the Q1D metal, that at high energy the Cooper channel and the
particle-hole channel are coupled, the usual ladder summation is not
adequate, and it is necessary to use the parquette approximation. Failure
of the ladder approximation implies the failure of the mean field theory,
since the two approaches are equivalent.

Fortunately, this coupling between the channels is a 1D feature, which
disappears at sufficiently low energy. Indeed, it is also proven in
\cite{prigodin_firsov} that the weakly interacting Q1D Fermi system
experience the dimensional crossover at low energy; below the crossover the
channels decouple, and the ladder approximation (hence, the mean field
theory) is valid again.

Thus, if we need to know a low-energy properties of the model (e.g., the
phase diagram), it is enough to derive the effective Hamiltonian valid below
the dimensional crossover, for this Hamiltonian may be analyzed with the
help of usual mean field approximation.

How does this Hamiltonian look like? This question is addressed in several
theoretical papers 
\cite{guay,mech,rozhkov,rozhkov_cond-mat}.
These papers discuss in detail the
dependence of the effective coupling constants on the bare one. However, for
our purposes it is enough to guess general features of the low-energy
Hamiltonian. Our conjecture is based on two assumptions about RG flow:

{\it (i)} at high energy the transverse single-electron hopping is the most
relevant operator in the problem, and the crossover occurs when the
effective transverse hopping becomes comparable to the running cutoff;

{\it (ii)} at high energy the SDW and CDW susceptibilities are the
dominant; among these two the SDW susceptibility prevails.

What do these statements mean physically? Assumption {\it (i)} is valid
provided that bare interactions are sufficiently weak (for an accurate 
criterion of the single-electron hopping relevance one can consult e.g.,
\cite{boson}).
It guarantees that our low-energy model is Fermi liquid with weak
effective coupling constants and low effective anisotropy:
\begin{eqnarray}
\tilde r = \tilde t / \tilde v_{\rm F} \tilde \Lambda \sim 1.
\label{r_eff}
\end{eqnarray}
When {\it (i)} is
violated, the system freezes into an ordered state (typically, SDW or CDW)
before reaching the Fermi liquid regime. Needless to say, our analysis is
inapplicable in such a situation.

Assumption {\it (ii)} is a consequence of the electron repulsion combined
with the fact that high-energy physics is purely one-dimensional. It is
known that the 1D metal has strongly divergent susceptibilities toward SDW
and CDW. Of these two, the former is stronger due to the
in-chain backscattering $g_1$. Because of all this,
the SDW susceptibility prevails in the high-energy regime.

This effect has nothing to do with the nesting properties of the actual Fermi
surface, which is a low-energy feature. Moreover, one can say that this
abundance of the high-energy modes, enhancing SDW correlations regardless of
the nesting, is a peculiarity of Q1D metal, which makes its physics so unusual.

Taking {\it (i)} and {\it (ii)} into account we can write the following
effective Hamiltonian:
\begin{eqnarray}
\tilde H 
= 
\sum_i 
	\tilde H^{\rm kin}_i + \tilde H^{\rho\rho}_i 
+
\sum_{ij}
	\tilde H^{\rm hop}_{ij} 
	+ 
	\tilde H^{\rho\rho}_{ij} 
	+ 
	\tilde H^{SS}_{ij}.
\label{Heff}
\end{eqnarray}
The spin-spin transverse interaction term
$\tilde H^{SS}_{ij}$
is equal to:
\begin{eqnarray}
\tilde H^{SS}_{ij}
&=&
\int dx
\Big[
	\tilde J_0^\perp (i-j) 
	{\bf S}_{ i} \cdot {\bf S}_{ j} 
	+ 
\label{HSS}
\\
	&&\tilde J_{2k_{\rm F}}^\perp (i-j) 
	\left( 
		{\bf S}_{2k_{\rm F} i} 
		\cdot
		{\bf S}_{-2k_{\rm F} j} 
         	+ 
		\text{H.c.} 
	\right)
\Big],
\nonumber 
\\
{\bf S}
& = &
\sum_{p \sigma \sigma'} \vec{\tau}_{\sigma \sigma'} 
\colon
	\psi^{\dagger}_{p \sigma }
	\psi^{\vphantom{\dagger}}_{p \sigma' } 
\colon.
\end{eqnarray}
This term, although absent in the microscopic Hamiltonian, appears at low
energies. 

Other terms of Eq. (\ref{Heff}), 
$\tilde H^{\rm kin}_i$,
$\tilde H^{\rho\rho}_i$, 
$\tilde H^{\rm hop}_{ij}$, 
and
$\tilde H^{\rho\rho}_{ij}$, 
have the same structure as the corresponding operators without tilde 
($H^{\rm kin}_i$,
$H^{\rho\rho}_i$, 
$H^{\rm hop}_{ij}$, 
and
$H^{\rho\rho}_{ij}$),
but the former have renormalized constants 
($\tilde v_{\rm F}$ instead of $v_{\rm F}$, $\tilde t$ instead of $t$,
$\tilde g$'s instead of $g$'s). On top of this, the cutoff of the effective
theory is much smaller than the microscopic cutoff:
$
\tilde \Lambda \ll \Lambda.
$

Hierarchy of the effective coupling constants differs from Eq. (\ref{ratio})
and Eq. (\ref{2kF<0}). At the dimensional
crossover the transverse hopping becomes comparable to the cutoff [see
Eq.(\ref{r_eff})].
The effective system remains anisotropic (for example, its Fermi surface
consists of two warped sheets disconnected from each other), yet, this
anisotropy is not as strong as the anisotropy of the original microscopic
system.

Since at high energy the
dominating fluctuations are SDW and CDW, $2k_{\rm F}$ coupling constants
are enhanced:
\begin{eqnarray}
\tilde g_{{2 k_{\rm F}}} \gg \tilde g_4,
\\
\tilde J_{{2 k_{\rm F}}} \gg \tilde g_4,
\\
\tilde J^\perp_{2k_{\rm F}} \gg \tilde J^\perp_0,
\label{eff_Jhier}
\\
\tilde g^\perp_{2k_{\rm F}} \gg \tilde g^\perp_0.
\label{eff_ghier}
\end{eqnarray}
All coupling constants are smaller than the renormalized Fermi velocity
$\tilde v_{\rm F}$.

It is tempting to declare that, since SDW correlations dominate over CDW
correlations in the high-energy regime, the SDW coupling constant
$\tilde J^\perp_{{2 k_{\rm F}}}$ 
is bigger than the CDW constant
$\tilde g^\perp_{{2 k_{\rm F}}}$.
However, this is not necessary
true for bare 
$J^\perp_{{2 k_{\rm F}}}$ 
is zero, while
$g^\perp_{{2 k_{\rm F}}} \ne 0$.
This might affect the outcome at the crossover scale.

\section{Phase diagram}
\label{phase_diag}

In this section we apply the mean-field analysis to the effective Hamiltonian 
$\tilde H$,
Eq. (\ref{Heff}). 

\subsection{Density wave phases}

The low temperature phase of the effective Hamiltonian depends on the
nesting properties of the Fermi surface. Assume first that only
nearest-neighbor hopping amplitude
$t_1$
is non-zero. In this case the Fermi surface nests perfectly. The SDW
susceptibility is equal to:
\begin{eqnarray}
\chi_{\rm SDW}   
=
\frac{1}{\pi \tilde v_{\rm F}}
         \ln\left( 
	 		\frac{2 \tilde v_{\rm F} \tilde \Lambda}{T}
	    \right).
\end{eqnarray}
The CDW susceptibility is the same.

As it is obvious from Eq. (\ref{HSS}), the coupling constant for SDW is
equal to 
$\tilde g_{\rm SDW} = \tilde J_{2 k_{\rm F}} + z \tilde J_{2k_{\rm F}}^\perp$,
where $z$ is the number of the nearest neighbours.
The usual mean-field equation for the critical temperature 
$\tilde g_{\rm SDW} \chi_{\rm SDW} (T_{\rm SDW}) = 1$
gives us the formula for $T_{\rm SDW}$:
\begin{eqnarray}
T_{\rm SDW}^{\rm max} 
= 
2 \tilde v_{\rm F} \tilde \Lambda 
\exp [ - \pi \tilde v_{\rm F}
	/ 
	(\tilde J_{2 k_{\rm F}} + z \tilde J_{2k_{\rm F}}^\perp) ].
\end{eqnarray}
The subscript `max' is to remind us that at perfect nesting the
transition temperature is the highest.

The CDW coupling constant 
$\tilde g_{\rm CDW} = \tilde g_{2 k_{\rm F}} + z \tilde g_{2k_{\rm F}}^\perp$
may be larger or smaller than
$\tilde g_{\rm SDW}$,
depending on the bare values of $g$, $g^\perp$, and
$J_{2k_{\rm F}}$.
The density wave type is determined by comparison of the coupling constants:
if
\begin{eqnarray}
\tilde g_{\rm CDW}
=
\tilde g_{2 k_{\rm F}} + z \tilde g_{2k_{\rm F}}^\perp
>
\tilde g_{\rm SDW}
=
\tilde J_{2 k_{\rm F}} + z \tilde J_{2k_{\rm F}}^\perp,
\label{sdw_vs_cdw}
\end{eqnarray} 
the ground state is CDW, otherwise, it is SDW. 

When the nesting is spoiled (for example, by introducing next-to-nearest
neighbor hopping amplitude $t_2$), the density wave critical temperature
decreases.  This happens because antinesting destroys
the divergence of the susceptibility. For example, one might write for SDW
(the case of CDW is identical):
\begin{eqnarray}
\chi_{\rm SDW}   \propto 
\frac{1}{\pi \tilde v_{\rm F}}
\times 
\cases{
         \ln\left( 2 \tilde v_{\rm F}\tilde \Lambda/T \right),
                  & if $T> \tilde t_2$,\cr
         \ln\left( 2 \tilde v_{\rm F} \tilde \Lambda/ \tilde t_2\right),
                  & if $T<\tilde t_2$,
      }
\label{chi}
\end{eqnarray} 
where 
$\tilde t_2$
is the renormalized value of $t_2$.

When $\tilde t_2$ is bigger than certain critical value:
\begin{eqnarray}
\tilde t_2 > t_2^c = T_{\rm SDW}^{\rm max},
\end{eqnarray}
the mean-field equation 
$\tilde g_{\rm SDW} \chi_{\rm SDW} (T_{\rm SDW}) = 1$
has no solution. Thus, exponentially small 
$\tilde t_2$
is enough to destroy the density wave phase.

\subsection{Superconductivity} 

When the antinesting destroys the density wave, the system becomes
superconducting. To demonstrate this let us introduce the following set of
Cooper pair creation operators:
\begin{eqnarray}
\label{sc_matrix}
(\hat \Delta_{ij})_{\sigma\sigma'}
=
\psi^\dagger_{{\rm L}\sigma i}
\psi^\dagger_{{\rm R}\sigma' j}.
\end{eqnarray}
Operator
$(\hat \Delta_{ij})_{\sigma\sigma'}$
creates a Cooper pair composed of a left-moving electron of spin $\sigma$ on
chain $i$ and of a right-moving electron of spin $\sigma'$ on chain $j$.

Matrix $\hat \Delta_{ij}$ may be symmetrized with respect to the chain
indices as well:
\begin{eqnarray}
\label{sc_symm}
\hat \Delta^{s/a}_{ij}
=
\frac{1}{2}
\left(
          \hat \Delta_{ij}
          \pm
          \hat \Delta_{ji}
\right).
\end{eqnarray}
The superscript `s' (`a') stands for `symmetric' (`antisymmetric').

Further, it is convenient to write
$\hat \Delta_{ij}^{s/a}$
as a sum of three symmetric matrices
${ i}\vec{\tau}\tau^y$
and one antisymmetric matrix ${i} \tau^y$:
\begin{eqnarray}
\label{sc_order}
\hat \Delta_{ij}^{s/a}
=
\frac{1}{\sqrt{2}}
\left[
           {\bf d}_{ij}^{s/a}
           \cdot
           ({i}\vec{\tau}\tau^y)
           +
           \Delta_{ij}^{s/a}
           {i} \tau^y
\right],
\end{eqnarray}
where
$\vec{\tau} = ( \tau^x, \tau^y, \tau^z)$
is a vector composed of three Pauli matrices. Operator
$\Delta_{ij}^{s/a}$
(${\bf d}_{ij}^{s/a}$)
creates a Cooper pair in a singlet (triplet) state.

Using these operators we can rewrite 
$\tilde H^{\rho \rho}_{ij}$
and
$\tilde H^{SS}_{ij}$:
\begin{eqnarray}
\label{sc_coupling}
\sum_{ij}    \tilde {H}^{\rho\rho}_{ij} + \tilde {H}^{SS}_{ij}
=
- \sum_{ij}
\int dx
\Big[
\tilde g_{x^2 - y^2}
        \Delta_{ij}^{s}
        (\Delta_{ij}^s)^\dagger
\qquad
\\
\nonumber
+
\tilde g_{xy} 
        \Delta_{ij}^{a}
        (\Delta_{ij}^a)^\dagger
+
\tilde g_f
\mathbf{d}_{ij}^s
\cdot
(\mathbf{d}_{ij}^s)^{\dagger}
+
\tilde g_f'
        \mathbf{d}_{ij}^a
        \cdot
        (\mathbf{d}_{ij}^a)^{\dagger}
\Big]
+ \ldots ,
\end{eqnarray}
where the ellipsis stand for the terms, which cannot be expressed as a
product of a Cooper-pair creation and a Cooper-pair destruction operators
(for example,
$\psi_{{\rm L} \sigma}^\dagger \psi_{{\rm L} \sigma'}^{\vphantom{\dagger}}
\psi_{{\rm L} \sigma''}^\dagger \psi_{{\rm L} \sigma'''}^{\vphantom{\dagger}}
$).
The coupling constants are:
\begin{eqnarray} 
\tilde g_{x^2 - y^2}
=
6 \tilde J^\perp_{2k_{\rm F}} 
-
2 \tilde g_{2 k_{\rm F}}^\perp
+
6 \tilde J^\perp_0
-
2\tilde g_0^\perp,
\\
\tilde g_{xy} 
=
- 6 \tilde J_{2k_{\rm F}}^\perp
+ 
2\tilde g_{2 k_{\rm F}}^\perp
+
6 \tilde J^\perp_0
-
2\tilde g_0^\perp,
\\
\tilde g_f
=
2\tilde J_{2k_{\rm F}}^\perp
+ 
2\tilde g_{2 k_{\rm F}}^\perp
-
2\tilde J_0^\perp
-
2\tilde g_0^\perp,  
\\
\tilde g_f'
=
- 2\tilde J^\perp_{2k_{\rm F}}
- 
2\tilde g_{2 k_{\rm F}}^\perp 
- 
2\tilde J_0^\perp
-
2\tilde g_0^\perp.
\label{gf'}
\end{eqnarray}

We see from Eq.~(\ref{gf'}) that the order parameter 
$\langle \mathbf{d}^a \rangle$
is always zero, since the coupling constant
$\tilde g_f'$ 
is always negative.

Three other order parameters may be non-zero. Consider first
$d_{x^2 - y^2}$-wave
($\langle \Delta^s \rangle \ne 0$).
This type of superconductivity is at least metastable, if:
\begin{eqnarray}
\tilde g_{x^2 - y^2} > 0
\Leftrightarrow
3 \tilde J^\perp_{2k_{\rm F}} 
>
\tilde g_{2 k_{\rm F}}^\perp.
\label{meta_x2y2}
\end{eqnarray}
In the latter inequality we neglected 
$\tilde g_0^\perp$
and
$\tilde J_0^\perp$
for they are small [see Eq.~(\ref{eff_Jhier}) and Eq.~(\ref{eff_ghier})].

Triplet $f$-wave superconductivity 
($\langle \mathbf{d}^s \rangle \ne 0$)
is always metastable, since
$\tilde g_f > 0$
[provided that Eq.~(\ref{eff_Jhier}) and Eq.~(\ref{eff_ghier}) are satisfied].
This guarantees that, after the density wave is destroyed by the
antinesting, the Q1D metal becomes a superconductor. 

Singlet $d_{xy}$-wave superconductivity 
($\langle \Delta^a \rangle \ne 0$)
is metastable, if
$\tilde g_{xy} > 0$,
which is equivalent to:
\begin{eqnarray}
\tilde g^\perp_{2k_{\rm F}} 
>
3 \tilde J_{2 k_{\rm F}}^\perp.
\end{eqnarray}

The true, stable ground state is determined by comparison of the mean-field
transition temperatures for different superconducting order parameters.
These temperatures are the solutions of the equations 
$g_\alpha \chi_\alpha (T^\alpha_c) = 1$, where
$\alpha$ is either  $f$, or $d_{x^2 - y^2}$, or $d_{xy}$.

Let us compare first
$T_c^{x^2 - y^2}$
and
$T_c^f$.
These two order parameters have identical orbital structure. Therefore,
their susceptibilities are the same:
$\chi_{x^2 - y^2} = \chi_f$.
Consequently, in order to determine the relative stability
of $d_{x^2-y^2}$-wave and $f$-wave we must compare 
$\tilde g_{x^2 - y^2}$ and $\tilde g_f$.
Specifically, 
$T^{x^2 - y^2}_c > T^f_c$, if:
\begin{eqnarray}
\tilde g_{2k_{\rm F}}^\perp <  \tilde J_{2k_{\rm F}}^\perp.
\label{f_vs_d}
\end{eqnarray}
Eq. (\ref{f_vs_d}) implies that spin-density fluctuations, which enhance
$\tilde J_{2k_{\rm F}}^\perp$,
favor $d_{x^2 - y^2}$-wave over $f$-wave
\cite{nickelII,flex}.
Thus, proximity to the SDW phase promotes the former type of order. On the
contrary, close to CDW the charge-density fluctuations intensify, effective
coupling 
$\tilde g_{2k_{\rm F}}^\perp$
grows, advancing the $f$-wave superconductivity. 

To make this argument more concrete, consider the transition separating CDW
and superconductivity.  Stability of CDW implies that
Eq.~(\ref{sdw_vs_cdw})
is fulfilled. This inequality may be rewritten as
\begin{eqnarray}
\tilde g_{2k_{\rm F}} - \tilde J_{2k_{\rm F}}
>
\frac{z}{4}
\left(
	\tilde g_{x^2 - y^2} - \tilde g_f
\right).
\end{eqnarray}
Observe that the bare constants satisfy [see Eq.~(\ref{g2kF}) and
Eq.~(\ref{J2kF})]:
\begin{eqnarray}
g_{2k_{\rm F}} < J_{2k_{\rm F}}.
\label{g<J}
\end{eqnarray}
If we assume that the renormalized coupling constants 
$\tilde g_{2k_{\rm F}}$
and
$\tilde J_{2k_{\rm F}}$
satisfy the same inequality:
\begin{eqnarray}
\tilde g_{2k_{\rm F}} < \tilde J_{2k_{\rm F}},
\label{g<J_eff}
\end{eqnarray} 
then we obtain:
\begin{eqnarray}
\tilde g_{x^2 - y^2} - \tilde g_f
< 0.
\end{eqnarray}
This means that {\it CDW cannot have common boundary with 
$d_{x^2 - y^2}$-wave superconductivity}.

However, inequality Eq.~(\ref{g<J}) is not equivalent to
Eq.~(\ref{g<J_eff}). Therefore, this argumentation points to a trend rather
than establishes a hard connection between the density wave type and the
order parameter symmetry.

The coupling constant for $d_{xy}$-wave is always smaller than
$\tilde g_f$:
\begin{eqnarray}
\tilde g_f - \tilde g_{xy} = 8 \tilde J_{2k_{\rm F}}^\perp > 0.
\label{diff}
\end{eqnarray} 
This does not necessary mean that $f$-wave superconductivity always
overshadows the $d_{xy}$-wave superconductivity. These two have different
orbital structure (the former is symmetric with respect to inversion of the
transverse coordinates, while the latter is antisymmetric). Thus,
it is possible to choose the density of states in such a way that
$\chi_{xy} > \chi_f$.
If this happens, and if the difference between 
$\tilde g_f$
and
$\tilde g_{xy}$
[see Eq.~\ref{diff}] is not too large, the $d_{xy}$-wave order parameter may
be more stable than $f$-wave. Therefore, we conclude that, when
$d_{x^2 - y^2}$-wave is
unstable, both $f$-wave and $d_{xy}$-wave can be possible choices for the
symmetry of the order parameter.

It is also important to note that the superconductivity becomes possible
only in a system with sufficiently pronounced anisotropy ($r \ll 1$).
Otherwise, the 1D renormalization is weak, and 
\begin{eqnarray}
\tilde g_{2k_{\rm F}}^\perp \approx g_{2k_{\rm F}}^\perp,
\quad
\tilde g_{0}^\perp \approx g_{0}^\perp,
\quad
\tilde J_{0, 2k_{\rm F}}^\perp \approx 0.
\end{eqnarray} 
As a result, instead of Eq.~(\ref{eff_Jhier}) and
Eq.~(\ref{eff_ghier}), one has to use Eq.~(\ref{2kF<0}). In this case 
all superconducting coupling constants are negative because they are
dominated by
$-2 \tilde g_0^\perp$
term.

\begin{figure} [!b]
\centering
\leavevmode
\epsfxsize=8cm
\epsfbox {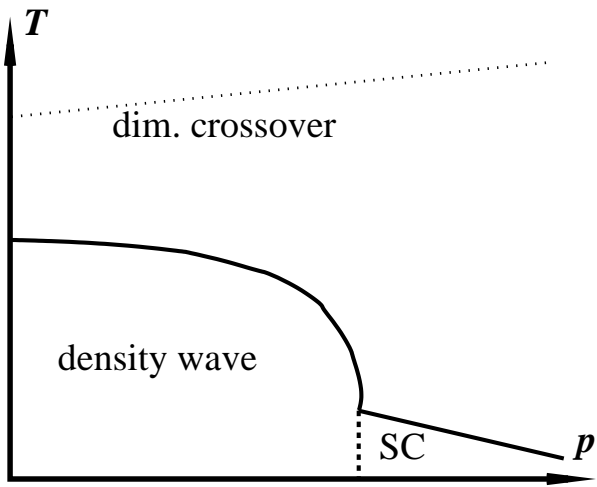}
\caption[]
{\label{fig1} 
Qualitative phase diagram of our model on the plane pressure-temperature.
Growth of the external pressure acts to increase the transverse hopping,
which decreases
the anisotropy and increases the antinesting. Solid lines show second-order
phase transitions into the density wave state (SDW or CDW) and the
superconducting phase. Dashed line shows the first-order transition between
the density wave and the superconductivity. The dotted line at high
temperature marks location of the dimensional crossover.
}
\end{figure}

On Fig.~\ref{fig1} we present the phase diagram, which emerges from our
discussion. It is drawn on the temperature-pressure plane. The effect of
the pressure is twofold. First, it increases the transverse hopping, which
leads to increase of the anisotropy ratio $r$. As a result, the dimensional
crossover temperature grows, and the superconducting critical temperature
decreases.

Second, the pressure increases the next-to-nearest neighbor hopping,
spoiling the nesting properties of the Fermi surface. Thus, the density
wave transition temperature vanishes, when the pressure exceeds some
critical value $p_c$. This explains the major features of our phase
diagram.

\section{Discussion}
\label{discussion}
We demonstrated under rather general assumptions that {\it (i)} Q1D metal
has a superconducting ground state, and
{\it (ii)} the symmetry of the order parameter is sensitive to the details
of the interaction and the density of states.

This implies that the ``universal'' phase diagram of the Bechgaard salts
\cite{universal} is a robust feature of Q1D metals, easily reproducible
theoretically. At the same time, prediction of the order parameter
symmetry on the basis of the microscopic model is virtually impossible for
it requires very accurate calculations of the competing states' energies.

One must remember that different Q1D organic superconductors, despite many
similarities they share, may have different symmetry of the superconducting
order parameter. Furthermore, it is possible that an individual sample
experience a phase transition between different types of superconductivity,
if the external parameters (pressure, magnetic field) change. It is
possible that such a phenomena is indeed observed experimentally:
within the superconducting region of (TMTSF)ClO a phase transition is
detected when the external magnetic field exceeds some critical value
\cite{knight}.

Our approach, however, has some limitations. Namely, it is not applicable
in situations, where the dimensional crossover occurs due to transverse
interaction rather than transverse hopping \cite{caron}, are beyond its
reach.

To conclude we discussed the symmetry of the order parameter in Q1D metallic
Fermi system. It is demonstrated that for a certain class of Q1D
superconductors the order parameter symmetry is a non-universal feature. It
is determined by a delicate interplay of microscopic constants
characterizing the system. The order parameter type may easily change in
response to variation of external (pressure) or internal (doping) factors.

\section{Acknowledgements}

The author is grateful for the support provided by the RFBR grants
No.~08-02-00212, and 09-02-00248.

\begin{thebibliography}{99}

\bibitem{guay} B. Guay and C. Bourbonnais, Synth. Metals {\bf 103}, (1999)
2180.

\bibitem{mech} C. Bourbonnais and L.G. Caron, Europhys. Lett., {\bf 5}, 209
(1988).

\bibitem{rozhkov} A.V. Rozhkov, { Phys. Rev. B}, {\bf 68}, 115108
(2003).

\bibitem{rozhkov_cond-mat} A.V. Rozhkov, arXive:0708.1884 (unpublished).

\bibitem{nickelII} J. C. Nickel, R. Duprat, C. Bourbonnais, and N. Dupuis,
{ Phys. Rev. B}, {\bf 73}, 165126 (2006).

\bibitem{dup} Raphael Duprat, C. Bourbonnais, { Eur. Phys. J. B},
{\bf 21}, 219 (2001).

\bibitem{nickel} J.C. Nickel, R. Duprat, C. Bourbonnais, and N. Dupuis,
{ Phys. Rev. Lett. }, {\bf 95}, 247001 (2005) and 
{ cond-mat/0502614}, v.2.

\bibitem{review_RG1} N. Dupuis, C. Bourbonnais and J.C. Nickel, 
{ Low Temp. Phys.}, {\bf 32}, 380 (2006).

\bibitem{review_RG2} C. Bourbonnais, preprint { cond-mat/0204345}.

\bibitem{fuseyaI} T. Aonuma, {\it et.al.}, arXiv:0806.2219 (unpublished).

\bibitem{review_RPA} Kazuhiko Kuroki, { J. Phys. Soc. Jpn.}, {\bf 75},
051013 (2006).

\bibitem{fuseyaII} Y. Fuseya and Y. Suzumura, J. Phys. Soc. Jpn., {\bf 74},
1263 (2005).

\bibitem{tanaka} Y.Tanaka and K. Kuroki, { Phys. Rev. B}, {\bf 70},
060502(R) (2004).

\bibitem{kuroki} Kazuhiko Kuroki and Yukio Tanaka, { J. Phys. Soc.
Jpn.}, {\bf 74}, 1694 (2005).

\bibitem{flex} Kazuhiko Kuroki, Ryotaro Arita, and Hideo Aoki, { Phys.
Rev. B}, {\bf 63}, 094509 (2001).

\bibitem{belmechri} N. Belmechri, {\it et.al.}, Europhys. Lett., {\bf 80},
37004 (2007).

\bibitem{belmechriII} N. Belmechri, {\it et.al.}, Europhys. Lett., {\bf 82},
47009 (2008).


\bibitem{NMR} I.J. Lee, S.E. Brown, W.G. Clark, M.J. Strouse, M.J. Naughton,
W. Kang, and P.M. Chaikin, { Phys. Rev. Lett.}, {\bf 88}, 017004 (2001).

\bibitem{NMR2} I.J. Lee, D.S. Chow, W.G. Clark, M.J. Strouse, M.J. Naughton,
P.M. Chaikin, and S.E. Brown, { Phys. Rev. B}, {\bf 68}, 092510 (2003).

\bibitem{imp} N. Joo, P. Auban-Senzier, C.R. Pasquier, P. Monod,
D. J\'erome, and K. Bechgaard, { Eur. Phys. J. B}, {\bf 40}, 43 (2004).

\bibitem{field} I.J. Lee and M.J. Naughton, G.M. Danner and P.M. Chaikin,
{ Phys. Rev. Lett.}, {\bf 78}, 3555 (1997).

\bibitem{field2} J.I. Oh and M.J. Naughton, { Phys. Rev. Lett.}, {\bf 92},
067001, (2004).

\bibitem{no_nodes} St\'ephane Belin and Kamran Behnia, { Phys. Rev.
Lett.}, {\bf 79}, 2125 (1997).



\bibitem{knight} J. Shinagawa, {\it et.al.}, Phys. Rev. Lett., {\bf 98},
147002 (2007).

\bibitem{universal} H. Wilhelm, D. Jaccard, R. Duprat, C. Bourbonnais,
D. Jerome, J. Moser, C. Carcel, and J. M. Fabre, { Eur. Phys. J. B},
{\bf 21}, 175 (2001).

\bibitem{prigodin_firsov} V.N. Prigodin and Yu.A. Firsov, Sov. Phys. JETP,
{\bf 49}, 369 (1979).

\bibitem{klutt} W. Kohn and J.M. Luttinger, {Phys. Rev. Lett.} {\bf 15},
524, (1965). 

\bibitem{boson} A.O. Gogolin, A.A. Nersesyan, A.M. Tsvelik, {
Bosonization and Strongly Correlated Systems}, (Cambridge University Press,
Cambridge, England, 1998).


\bibitem{caron} L. Caron and C. Bourbonnais, Physica {\bf 143B}, 453 (1986).

\end{thebibliography}
\end{document}